\documentclass[%
 reprint,
 amsmath,amssymb,
 aps,
]{revtex4-2}

\usepackage{graphicx}
\usepackage{dcolumn}
\usepackage{bm}
\usepackage{color}

\begin{document}

\preprint{APS/123-QED}
\newcommand{\ket}[1]{\left\vert{#1}\right\rangle}
\newcommand{\bra}[1]{\left\langle{#1}\right\vert}
\newcommand{\abs}[1]{\left\vert{#1}\right\vert}
\title{Correlated two-photon scattering in a one-dimensional waveguide coupled to two- or three-level giant atoms}

\author{Wenju Gu}
 \email{guwenju@yangtzeu.edu.cn}
 \altaffiliation{School of Physics and Optoelectronic Engineering, Yangtze University, Jingzhou 434023, China}
\author{He Huang}%
\affiliation{School of Physics and Optoelectronic Engineering, Yangtze University, Jingzhou 434023, China}%
\author{Zhen Yi}%
\affiliation{School of Physics and Optoelectronic Engineering, Yangtze University, Jingzhou 434023, China}
\author{Lei Chen}%
\affiliation{School of Physics and Optoelectronic Engineering, Yangtze University, Jingzhou 434023, China}
\author{Lihui Sun}%
\affiliation{School of Physics and Optoelectronic Engineering, Yangtze University, Jingzhou 434023, China}

\author{Huatang Tan}
 \email{tht@mail.ccnu.edu.cn}
\affiliation{Department of Physics, Huazhong Normal University, Wuhan 430079, China}%

\date{\today}

\begin{abstract}
We investigate the two-photon scattering processes in a one-dimensional waveguide coupled to either a two-level or three-level giant atom. By manipulating the accumulated phase shift between the two coupling points, we are able to effectively modify the characteristics of these scattering processes. Utilizing the Lippmann-Schwinger formalism, we obtain the exact two-photon interacting scattering wavefunctions of these two systems. Additionally, analytical expressions for the incoherent power spectra and second-order correlations are derived. The incoherent spectrum, which is defined by the correlation of the bound state, provides valuable insights into photon-photon correlations. It serves as a useful indicator of the degree of photon-photon correlation between scattered photons. Furthermore, the second-order correlation function gives a direct measure of the photon-photon correlation. For photons scattered by the two-level giant atom, manipulating the accumulated phase shift allows for improvement of the photon-photon correlation and adjustment of the evolution of the second-order correlation. In the case of the three-level giant atom, the photon-photon correlation can be substantially increased. The photon-photon interaction and correlation distance of the scattered photons can be further enhanced by tuning the accumulated phase shift. Moreover, the statistical properties can be adjusted by the control field.
\end{abstract}

\maketitle


\section{Introduction}
Exploiting atom-photon interactions plays a critical role in the emerging field of waveguide quantum electrodynamics (QED)~\cite{D.Roy@rmp2017,X.Gu@PR2017,A.S.Sheremet@rmp2023,P.Forn-Diaz@nphy2017}, which opens up novel opportunities for both fundamental physics~\cite{D.E.Chang@natphoton2014,T.Shi@njp2018,N.Fayard@PRR2021} and quantum information processing~\cite{B.Kannan@natphys2023,A.Gonzalez@prl2015,P.Forn-Diaz@prapp2017}. The conventional atom-photon interaction has been studied within the small atom regime, where the interaction is localized due to the presence of atoms in a region significantly smaller than the wavelength of photons. However, recent investigations have focused on a new paradigm known as ``giant atoms'', which exhibits nonlocal couplings. This has been achieved by coupling artificial atoms to propagating fields (e.g., surface acoustic waves) with wavelengths smaller than atomic sizes~\cite{G.Andersson@nphys2019}, or through meandering waveguides at separated points~\cite{B.Kannan@nat2020,A.M.Vadiraj@pra2021}. The introduction of nonlocal interactions enables the occurrence of various remarkable phenomena that are not achievable in the case of small atoms. These include frequency-dependent Lamb shifts and relaxation rates~\cite{A.F.Kockum@pra2014}, waveguide-mediated decoherence-free subspaces~\cite{A.F.Kockum@prl2018,A.Carollo@PRR2020}, nonexponential decay~\cite{G.Andersson@nphys2019, S.Longhi@ol2020, S.Guo@pra2020,Qiu2023}, oscillating bound states~\cite{L.Guo@PRR2020}, and the enhanced spontaneous sudden birth of entanglement~\cite{PhysRevLett.130.053601}.

In waveguide QED systems, the presence of a one-dimensional (1D) continuum of modes allows for a broader range of field bandwidths and facilitates strong coupling to local emitters by reducing the mode volume~\cite{D.Roy@rmp2017}. The study of few-photon scattering is well-established in the strong-coupling regime, where the interaction between light and the emitter dominates over loss and dephasing effects. Various theoretical techniques have been developed to investigate few-photon scattering in the context of small atoms. These techniques include the real-space Bethe-ansatz method~\cite{J.T.Shen@ol2005,J.T.Shen@prl2005,PhysRevB.81.155117,PhysRevLett.106.053601,PhysRevA.87.063819}, the Lehmann-Symanzik-Zimmermann reduction~\cite{T.Shi@prb2009,T.Shi@pra2011}, the wave-packet evolution approach~\cite{J.Q.Liao@pra2010,J.Q.Liao@pra2013}, the Lippmann-Schwinger (LS) formalism~\cite{H.Zheng@prl2013,Y.Fang@EPJ2014,PhysRevA.83.043823}, input-output formalism~\cite{S.Fan@pra2010,E.Rephaeli@pra2011}, as well as the Dyson series summation~\cite{D.L.Hurst@pra2018}.

The phenomenon of photon scattering in waveguide QED systems involving giant atoms has gained significant attention due to the unique interference effects~\cite{W.Zhao@pra2020,S.L.Feng@pra2021,L.Du@PRR2021,X.L.Yin@pra2022,PhysRevA.106.033522, Chen2022,Zhao2022}. It is worth noting that the majority of studies in this area primarily focus on single-photon interactions, with relatively limited exploration of few-photon scattering within the giant atom regime. The presence of nonlinearity in atomic systems would induce photon-photon correlations, and such photon-photon correlations can, in turn, influence the transmission and reflection of photons through phenomena like induced tunneling or blocking of photons~\cite{H.Zheng@pra2012,Z.Yi@adp2023,W.Gu@pra2022,J.Tang@pra2020}. Therefore, gaining a clear understanding of the nature of photon-photon correlations is crucial to comprehending the functioning of various quantum devices, such as a switchable mirror~\cite{E.S.Redchenko@nc2023} or a single-photon router~\cite{I.C.Hoi@prl2012}. In the context of few-photon scattering, the wavefunctions typically exhibit a common structure: the two-particle plane wave with momenta of photons rearranged and the bound state. The plane-wave component arises from the coherent scattering, while the bound state component originates from the incoherent scattering. The bound state decays exponentially as the distance between the two photons increases, and is associated with the two-particle irreducible T-matrix in scattering theory~\cite{S.Xu@prl2013}. Recently, the eigenstates of the scattering matrix have been successfully obtained for the even mode in a two-level giant system using a reasonable wavefunction hypothesis~\cite{Cheng2023}.

In this paper, we employ the LS formalism to analyze the two-photon scattering processes involving both two- and three-level giant atoms coupled to a 1D waveguide. By utilizing this approach, we are able to obtain the analytical two-photon interacting scattering wavefunctions for these systems. Additionally, the incoherent power spectrum is derived from the correlation of the bound state, with the total flux serving as an indicator of photon-photon correlation. The second-order correlation function provides a direct measure of photon-photon correlation. Through our analysis, we find that the accumulated phase shift can be utilized to enhance the photon-photon correlation and control the evolution of the second-order correlation for photons scattered by the two-level giant atom. In comparison to the two-level giant atom, the photon-photon correlation can be substantially enhanced for the system of the three-level giant atom. Moreover, we find that tuning the accumulated phase shift between the two coupling points allows for further improvement of photon-photon interaction and an increase in the decay distance of the second-order correlation. These properties can be explained by the poles of these systems. Furthermore, the statistics property of the scattering photons in the three-level giant atom system can be controlled by adjusting the control laser at the resonance condition. In the presence of the control field, the incident photons pass by the system coherently, thereby maintaining unchanged statistics. However, in the absence of the control field, the system effectively behaves as a two-level model, resulting in bunched transmitted photons and antibunched reflected photons.

The paper is organized as follows. In Sec.~\ref{sec2},  we analyze the two-photon scattering process in the system of the two-level giant atom coupled to a 1D waveguide. We investigate various aspects, including the derivation of the two-photon interacting scattering eigenstate, the incoherent power spectrum, and the second-order correlation function. In Sec.~\ref{sec3}, we extend our analysis to the case of the three-level giant atom, where we examine similar quantities such as the two-photon interacting scattering eigenstate, the incoherent power spectrum, and the second-order correlation function. The conclusions drawn from our study are given in Sec.~\ref{conclusion}. Additionally, a summary of technical details relevant to the calculation of the incoherent power spectrum is presented in Appendix.

\section{The system of a two-level giant atom coupled to a 1D waveguide}
\label{sec2}
\begin{figure}[hbt]
\centering\includegraphics[width=8cm,keepaspectratio,clip]{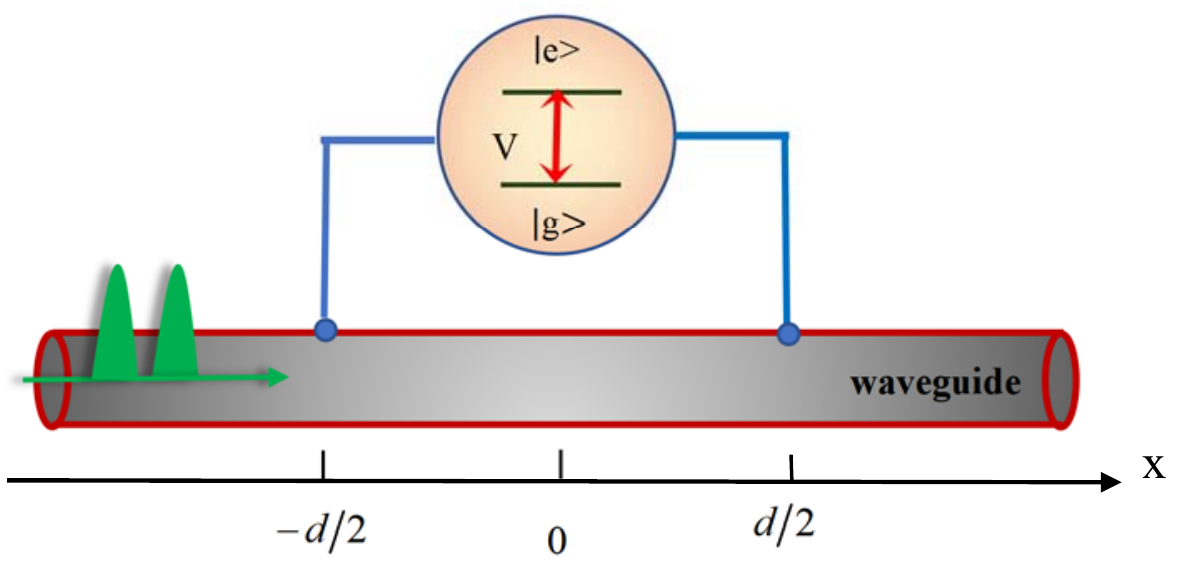}
\caption{Schematic illustration of a two-level giant atom side coupled to a one-dimensional (1D) waveguide. The atomic transition between the ground state $\ket{g}$ and the excited state $\ket{e}$ is coupled to waveguide modes positioned at $x=-d/2$ and $d/2$. The coupling strength is denoted as $V$.}
\label{fig1}
\end{figure}

We begin by considering the correlated two-photon dynamics in a system consisting of a two-level giant atom side coupled to a 1D waveguide, as shown in Fig.~\ref{fig1}. The configuration can be feasibly realized experimentally by establishing a coupling between a superconducting qubit and either meandering microwave transmission lines~\cite{A.M.Vadiraj@pra2021,B.Kannan@nat2020} or propagating surface acoustic waves (SAWs)~\cite{G.Andersson@nphys2019,F.Kockum}. In the microwave transmission line setup, the size of the qubit is significantly smaller than the wavelength, thereby enabling us to treat it as a point-like object at each coupling point. In the SAWs system, the interdigital transducer (IDT) forms the two transmon islands and couples with the SAWs propagating on the substrate. The entire IDT consists of two local IDTs at opposite ends, which are sufficiently distant from each other. Each IDT consists of $N$ pairs of fingers, arranged such that the distance between neighboring finger pairs matches the corresponding phonon wavelength, allowing for a constant coupling. Moreover, the distance between the centers of the two local IDTs can be made to be considerably larger than the SAW wavelengths, permitting us to approximate the interaction as point couplings. For convenience, let us assume that the atomic transition $\ket{g}\leftrightarrow\ket{e}$ interacts with the waveguide field at the coupling points of $x=-d/2$ and $d/2$, respectively. While the scattering process involving single photons has been extensively studied in previous works~\cite{Zhao2022,PhysRevA.104.033710}, our focus in this paper lies on the investigation of two-photon dynamics in order to explore photon-photon interactions. In real space, the Hamiltonian governing the system can be written as ($\hbar=1$ hereafter)
\begin{align}
\hat{H}=&\left(\omega_0-i\gamma_e/2\right)\hat{\sigma}_{ee}\nonumber\\&-i\nu_g\int dx\left[\hat{a}^\dag_R(x)\partial_x\hat{a}_R(x)-\hat{a}^\dag_L(x)\partial_x\hat{a}_L(x)\right]\nonumber\\
&+\frac{V}{2}\sum_{\alpha=R,L}\int dx M(x)\left[\hat{a}^\dag_\alpha(x)\hat{\sigma}^-+\hat{\sigma}^+\hat{a}_\alpha(x)\right].
\end{align}
Here $M(x)=\delta(x+d/2)+\delta(x-d/2)$ denotes the coupling points. The operators $\hat{a}^\dag_R(x)$ and $\hat{a}^\dag_L(x)$ are the creation operators for the right-moving and left-moving modes in real space, and $\nu_g$ is the group velocity. We will take $\nu_g=1$ for simplicity. For the two-level giant atom, $\omega_0$ is the atomic transition frequency, and $\gamma_e$ is the spontaneous emission rate of the excited state to modes other than the waveguide continuum. The atomic transition $\ket{g}\leftrightarrow\ket{e}$ couples to the waveguide modes at $x=-d/2$ and $x=d/2$ with an equal strength $V=\sqrt{2\Gamma}$, where $\Gamma$ denotes the atomic spontaneous decay rate to the waveguide continuum. In the strong coupling limit, the value of Purcell factor $P$ is large, i.e., $P=\Gamma/\gamma_e\gg1$.  For instance, $\Gamma/2\pi=2\text{MHz}$ and $\gamma_e/2\pi=0.03\text{MHz}$ are demonstrated in the experimental setup~\cite{B.Kannan@nat2020}. This indicates that the spontaneous decay is mainly to the waveguide compared with all the other modes. In this case, the atomic spontaneous dissipation rate $\gamma_e$ can be neglected.

The total excitation number is conserved for the Jaynes-Cummings model of interaction between the atom and waveguide field. Thus, in the single-excitation subspace, the eigenstate of the system can be written in the form
\begin{align}
\ket{\phi_1(k)}_\alpha=&\int dx\left[\phi_R^\alpha(k,x)\hat{a}^\dagger_R(x)+\phi_L^\alpha(k,x)\hat{a}^\dag_L(x)\right]\ket{0,g}\nonumber\\&+u_e^\alpha(k)\ket{0,e},
\end{align}
where $\alpha=\{R, L\}$, and $\phi_{R/L}^\alpha(k,x)$ denote the probability amplitudes of creating the right-moving and left-moving photons in real space for the incident photon with the wavevector $k$ in the $\alpha$-direction. Here $u_e^\alpha(k)$ is the excitation amplitude of the atom, and $\ket{0,g}$ denotes the vacuum state of the system. The probability amplitudes can be determined by the Schr\"{o}dinger equation $\hat{H}\ket{\phi_1(k)}_\alpha=k\ket{\phi_1(k)}_\alpha$, which gives
\begin{align}
\left(\omega_0-k\right)u_e^\alpha(k)+\sqrt{\frac{\Gamma}{2}}\sum_{\alpha^\prime=R,L}\int dx M(x)\phi_{\alpha^\prime}^\alpha(k,x)&=0,\nonumber\\
(-i\partial_x-k)\phi_R^\alpha(k,x)+\sqrt{\frac{\Gamma}{2}}M(x)u_e^\alpha(k)&=0,\nonumber\\
(i\partial_x-k)\phi_L^\alpha(k,x)+\sqrt{\frac{\Gamma}{2}}M(x)u_e^\alpha(k)&=0.
\end{align}
The solution of the wavefunction takes the following form
\begin{align}
\phi_R^R(k,x)&=\frac{e^{ikx}}{\sqrt{2\pi}}\big\{\theta(-d/2-x)+t_1(k)\big[\theta(x+d/2)\nonumber\\&-\theta(x-d/2)\big]+t_2(k)\theta(x-d/2)\big\},\nonumber\\
\phi_L^R(k,x)&=\frac{e^{-ikx}}{\sqrt{2\pi}}\big\{r_1(k)\theta(-d/2-x)\nonumber\\&+r_2(k)\left[\theta(x+d/2)-\theta(x-d/2)\right]\big\},\nonumber\\
\phi_R^L(k,x)&=\frac{e^{ikx}}{\sqrt{2\pi}}\big\{r_1(k)\theta(x-d/2)\nonumber\\&+r_2(k)\left[\theta(x+d/2)-\theta(x-d/2)\right]\big\},\nonumber\\
\phi_L^L(k,x)&=\frac{e^{-ikx}}{\sqrt{2\pi}}\big\{\theta(x-d/2)+t_1(k)\big[\theta(x+d/2)\nonumber\\&-\theta(x-d/2)\big]+t_2(k)\theta(-d/2-x)\big\},
\end{align}
where $\theta(x)$ is the Heaviside step function. Here, $t_2(k)$ and $r_1(k)$ denote the single-photon transmission and reflection amplitudes in the regions where $x>d/2$ and $x<-d/2$, respectively. Additionally, $t_1(k)$ and $r_2(k)$ are the probability amplitudes for transmission and reflection between two coupling points $-d/2<x<d/2$. These amplitudes are calculated as
\begin{align}
t_1(k)&=\frac{\omega_0-k-i\Gamma/2\left(1+e^{ikd}\right)}{\omega_0-k-i\Gamma\left(1+e^{ikd}\right)},\nonumber\\
t_2(k)&=\frac{\omega_0-k+\Gamma\sin kd}{\omega_0-k-i\Gamma\left(1+e^{ikd}\right)},\nonumber\\
r_1(k)&=\frac{i\Gamma\left(1+\cos kd\right)}{\omega_0-k-i\Gamma\left(1+e^{ikd}\right)},\nonumber\\
r_2(k)&=\frac{i\Gamma/2\left(1+e^{ikd}\right)}{\omega_0-k-i\Gamma\left(1+e^{ikd}\right)},\nonumber\\
u_e^\alpha(k)&=\frac{-\sqrt{\Gamma/\pi}\cos(kd/2)}{\omega_0-k-i\Gamma\left(1+e^{ikd}\right)}.
\end{align}

In order to study the two-photon scattering process, we employ the LS equation to obtain the eigenstate for two-photon scattering~\cite{H.Zheng@prl2013, Y.Fang@EPJ2014}. To account for the nonlinear effect of the two-level atom, we use a bosonic representation which includes an on-site interaction~\cite{P.Longo@prl2010,H.Zheng@prl2013},
\begin{align}
H=H_0+V,\hspace{5pt} V=\frac{U}{2}\hat{d}^\dagger\hat{d}(\hat{d}^\dagger\hat{d}-1).
\end{align}
The atomic operators $\hat{\sigma}^\pm$ in $\hat{H}_0$ are replaced by the bosonic creation and annihilation operators $\hat{d}^\dagger$ and $\hat{d}$, respectively. To map the atomic ground and excited states to zero- and one-boson states, $U\rightarrow\infty$ should be taken in the end to eliminate occupations greater than $1$. In this bosonic representation, when $U=0$, the Hamiltonian corresponds to a non-interacting Hamiltonian that can be easily solved. The non-interacting two-photon eigenstate is given by
\begin{align}
\ket{\phi_2(k_1,k_2)}_{\alpha_1\alpha_2}=\frac{1}{\sqrt{2}}\ket{\phi_1(k_1)}_{\alpha_1}\otimes\ket{\phi_1(k_2)}_{\alpha_2}.
\end{align}
The LS equation establishes a relationship between the fully interacting two-photon eigenstates $\ket{\psi_2(k_1,k_2)}_{\alpha_1\alpha_2}$ and $\ket{\phi_2(k_1,k_2)}_{\alpha_1\alpha_2}$ as
 \begin{align}
\ket{\psi_2(k_1,k_2)}_{\alpha_1\alpha_2}=&\ket{\phi_2(k_1,k_2)}_{\alpha_1\alpha_2}\nonumber\\&+G^R(E)V\ket{\psi_2(k_1,k_2)}_{\alpha_1\alpha_2},
 \label{eqn:Lip-Sch}
 \end{align}
 where $G^R(E)=1/(E-H_0+i0^+)$ is the retarded Green's function, and $E=k_1+k_2$ is the total energy of two incident photons. The two-particle identity operator in real space can be expressed as
 \begin{align}
 I_2=&I_2^x\otimes\ket{0}\bra{0}+I_1^x\otimes\ket{d}\bra{d}+I_0^x\otimes\ket{dd}\bra{dd},\nonumber\\
 I_n^x=&\sum_{\alpha_1\cdots\alpha_n}\int dx_1\cdots dx_n\ket{x_1\cdots x_n}_{\alpha_1\cdots\alpha_n}\bra{x_1\cdots x_n},
 \end{align}
where $\ket{0}$ is the ground state, and we denote the single-excitation state $\ket{d}=\hat{d}^\dagger\ket{0}$ and the two-excitation state $\ket{dd}=\hat{d}^{\dagger2}\ket{0}/\sqrt{2}$. By inserting the identity operator into Eq.~\eqref{eqn:Lip-Sch}, we have
 \begin{align}
&\ket{\psi_2(k_1,k_2)}_{\alpha_1\alpha_2}\nonumber\\&=\ket{\phi_2(k_1,k_2)}_{\alpha_1\alpha_2}+G^R(E)VI_2\ket{\psi_2(k_1,k_2)}_{\alpha_1\alpha_2}\nonumber\\
 &=\ket{\phi_2(k_1,k_2)}_{\alpha_1\alpha_2}+UG^R(E)\ket{dd}\langle{dd}\ket{\psi_2(k_1,k_2)}_{\alpha_1\alpha_2}.
 \label{eqn:L-S}
 \end{align}
To determine the two-photon interacting eigenstate, we project Eq.~\eqref{eqn:L-S} onto $\bra{dd}$:
\begin{align}
\langle{dd}\ket{\psi_2(k_1,k_2)}_{\alpha_1\alpha_2}=&\left(1-UG_{dd}\right)^{-1}\nonumber\\&\times\langle{dd}\ket{\phi_2(k_1,k_2)}_{\alpha_1\alpha_2},
\end{align}
where $G_{dd}=\bra{dd}G^R(E)\ket{dd}$. Next, we project Eq.~\eqref{eqn:L-S} onto a two-photon basis state $_{\alpha_1^\prime\alpha_2^\prime}\bra{x_1x_2}$ and taking the limit  $U\rightarrow\infty$, the two-photon interacting eigenstate eventually becomes
\begin{align}
&_{\alpha_1^\prime\alpha_2^\prime}\langle x_1x_2\ket{\psi_2(k_1,k_2)}_{\alpha_1\alpha_2}=_{\alpha_1^\prime\alpha_2^\prime}\langle x_1x_2\ket{\phi_2(k_1,k_2)}_{\alpha_1\alpha_2}\nonumber\\&-G_{xd}^{\alpha_1^\prime\alpha_2^\prime}(x_1,x_2)G_{dd}^{-1}\langle dd\ket{\phi_2(k_1,k_2)}_{\alpha_1\alpha_2},
\label{eqn:two-photon_wavefunction}
\end{align}
where $G_{xd}^{\alpha_1^\prime\alpha_2^\prime}(x_1,x_2)=_{\alpha_1^\prime\alpha_2^\prime}\bra{x_1x_2}G^R(E)\ket{dd}$, and $x_1$ and $x_2$ refer to the positions of the photons. The first term denotes the plane-wave state, while the second term contains all the nonlinearity and is commonly referred to as the two-photon bound state. In order to solve the Green's functions, we utilize the two-photon non-interacting scattering eigenstates to establish a two-particle identity in momentum space
\begin{align}
I_2^\prime=\sum_{\alpha_1,\alpha_2}\int dk_1dk_2\ket{\phi_2(k_1,k_2)}_{\alpha_1\alpha_2}\bra{\phi_2(k_1,k_2)}.
\end{align}
The Green's functions can be finally derived in the following form
\begin{widetext}
\begin{align}
G_{dd}&=\sum_{\alpha_1,\alpha_2}\int dk_1dk_2\frac{\langle dd\ket{\phi_2(k_1,k_2)}_{\alpha_1\alpha_2}\bra{\phi_2(k_1,k_2)}dd\rangle}{E-k_1-k_2+i0^+},\nonumber\\
G_{xd}^{\alpha_1\alpha_2}(x_1,x_2)&=\sum_{\alpha_1^\prime,\alpha_2^\prime}\int dk_1dk_2\frac{_{\alpha_1\alpha_2}\langle x_1x_2\ket{\phi_2(k_1,k_2)}_{\alpha_1^\prime\alpha_2^\prime}\bra{\phi_2(k_1,k_2)}dd\rangle}{E-k_1-k_2+i0^+},
\end{align}
\end{widetext}
with the elements
\begin{align}
\langle dd\ket{\phi_2(k_1,k_2)}_{\alpha_1\alpha_2}&=u_e^{\alpha_1}(k_1)u_e^{\alpha_2}(k_2),\nonumber\\
_{\alpha_1^\prime\alpha_2^\prime}\langle x_1x_2\ket{\phi_2(k_1,k_2)}_{\alpha_1\alpha_2}&=\frac{1}{2}\Big[\phi_{\alpha_1^\prime}^{\alpha_1}(k_1,x_1)\phi_{\alpha_2^\prime}^{\alpha_2}(k_2,x_2)\nonumber\\&+
\phi_{\alpha_1^\prime}^{\alpha_2}(k_2,x_1)\phi_{\alpha_2^\prime}^{\alpha_1}(k_1,x_2)\Big].
\end{align}

When the propagating time of photons between the coupling points is shorter than the atomic lifetime $1/\Gamma$, the Markovian approximation can be applied~\cite{PhysRevLett.113.183601}. In this case, the wavevector $k$ in the phase factors can be replaced by a constant $k_0=\omega_0/\nu_g$, and the accumulated phase shift between the coupling points is denoted as $\vartheta=k_0d$. The phase shift can be adjusted by changing the distance between these points. Then, by performing double integrals with the use of the standard contour integral techniques, the Green's functions become
\begin{align}
G_{dd}&=\frac{1}{E-2\omega_0+i2\Gamma^\prime},\nonumber\\
G_{xd}^{RR}(x_1,x_2)&=\frac{-\Gamma(1+\cos \vartheta)}{E-2\omega_0+i2\Gamma^\prime}e^{iEx_c}e^{i(E/2-\omega_0)\abs{x}-\Gamma^\prime\abs{x}},
\end{align}
where $\Gamma^\prime=\Gamma(1+e^{i\vartheta})$. During the contour integration, the conditions $x_1>d/2$, $x_2>d/2$, and $x=x_2-x_1$, $x_c=(x_1+x_2)/2$ are used. It can be proven that $G_{xd}^{LL}(-x_1,-x_2)=G_{xd}^{RR}(x_1,x_2)=G_{xd}^{RL}(x_1,-x_2)=G_{xd}^{LR}(-x_1,x_2)$ because of the parity symmetry. Substituting these expressions into Eq.~\eqref{eqn:two-photon_wavefunction}, the two-photon interacting eigenstate becomes
\begin{align}
\ket{\psi_2(k_1,k_2)}_{RR}=&\int dx_1dx_2\Bigg[\frac{f_{RR}(x_1,x_2)}{\sqrt{2}}\hat{a}^\dagger_R(x_1)\hat{a}^\dagger_R(x_2)\nonumber\\
&+\frac{f_{LL}(x_1,x_2)}{\sqrt{2}}\hat{a}^\dagger_L(x_1)\hat{a}^\dagger_L(x_2)\nonumber\\
&+f_{RL}(x_1,x_2)\hat{a}^\dagger_R(x_1)\hat{a}^\dagger_L(x_2)\Bigg]\ket{0}.
\end{align}
The coefficients are
\begin{align}
f_{RR}(x_1,x_2)=&\frac{e^{iEx_c}}{\sqrt{2}\pi}\left[t_2(k_1)t_2(k_2)\cos\Delta_1x+B_{k_1k_2}(x)\right],\nonumber\\
f_{LL}(x_1,x_2)=&\frac{e^{-iEx_c}}{\sqrt{2}\pi}\left[r_1(k_1)r_1(k_2)\cos\Delta_1x+B_{k_1k_2}(x)\right],\nonumber\\
f_{RL}(x_1,x_2)=&\frac{e^{iEx/2}}{2\pi}\Big[t_2(k_1)r_1(k_2)e^{2i\Delta_1x_c}\nonumber\\&+r_1(k_1)t_2(k_2)e^{-2i\Delta_1x_c}+2B_{k_1k_2}(x_c)\Big],\nonumber\\
B_{k_1k_2}(x)=&\frac{\Gamma^2(1+\cos\vartheta)^2e^{i(E/2-\omega_0)|x|-\Gamma^\prime|x|}}{\left(E/2-\omega_0+i\Gamma^\prime\right)^2-\Delta_1^2},
\end{align}
where $\Delta_1=(k_1-k_2)/2$ corresponds to half of the energy difference between two incident photons. For the case of $\vartheta=0$, the interacting eigenstate can be simplified to the same form as that of the small two-level atom scenario with a single coupling point in Ref.~\cite{J.T.Shen@pra2007} by replacing $\Gamma$ with $\Gamma/4$. The factor $4$ arises from the constructive interference between two coupling points.

For the input of the two-particle plane wave, the scattering wavefunction exhibits a common constituent: the two-particle plane wave with rearranged momenta of the photons and the bound state. The plane wave originates from the coherent scattering, while the bound state originates from the incoherent scattering. As the distance between the two photons increases, the bound state decays exponentially. The bound state can also be referred to as the two-particle irreducible T-matrix in scattering theory~\cite{S.Xu@prl2013}.

\subsection{Incoherent power spectrum}
The two-photon interacting eigenstate is composed of two components: the plane wave arising from coherent scattering, and the bound state arising from photon-photon interactions. To investigate their impacts on scattering processes, we first consider the power spectrum or resonance fluorescence, which is the Fourier transform of the first-order correlation function,
\begin{align}
S_\alpha(\omega)=\int dt e^{-i\omega t}\bra{\psi_2}\hat{a}^\dagger_\alpha(x_0)\hat{a}_\alpha(x_0+t)\ket{\psi_2},
\label{eqn:power_spectrum}
\end{align}
where $x_0$ is the position of the detector located far away from the scattering region. Here $S_\alpha(\omega)$ accounts for the spectral decomposition of the photons in the two-photon interacting wavefunction $\ket{\psi_2}$. The detailed derivation of the power spectrum is presented in Appendix A. In general, the power spectrum consists of the coherent and incoherent parts, i.e., $S_\alpha(\omega)=S_\alpha^{\text{coh}}(\omega)+S_\alpha^{\text{incoh}}(\omega)$. The coherent scattering component contributes a $\delta$-function, while the incoherent scattering component is determined by the correlation of the bound state in the wavefunction, as shown in Eq.~\eqref{eqn:incoherent_spectrum}. 

To explore the physical implications of the incoherent power spectrum, we consider the case where $\alpha=R$, specifically focusing on the two-photon transmission state. In this case, the incoherent power spectrum can be expressed in the form
\begin{align}
S_R^{\text{incoh}}(\omega)=\frac{4}{\pi^2}\abs{\Upsilon(\omega)}^2,
\end{align}
where
\begin{align}
\Upsilon(\omega)=&\frac{\Gamma^2(1+\cos\vartheta)^2}{(E/2-\omega_0+i\Gamma^\prime)}\nonumber\\&\times\frac{1}{(E-\omega_0-\omega+i\Gamma^\prime)(\omega-\omega_0+i\Gamma^\prime)}.
\end{align}
This expression is provided in Appendix A. When transforming the transmitted state from real space to frequency space, it can be represented as $t_2(\omega_1)t_2(\omega_2)\hat{a}_R^\dagger(\omega_1)\hat{a}_R^\dagger(\omega_2)\ket{0}+\frac{i}{\pi}\int d\omega\Upsilon(\omega)\hat{a}_R^\dagger(E-\omega)\hat{a}^\dagger_R(\omega)\ket{0}$.
The first term describes the independent propagation of the two photons, while the second term represents the final state of the two photons after undergoing inelastic scattering. According to the principle of energy conservation, the two scattered photons are always generated in pairs with frequencies of opposite signs. The coefficient $\Upsilon(\omega)$ quantifies the generation of these photon pairs. Therefore, the incoherent power spectrum serves as a direct measure of the production of photon pairs with a frequency of $\omega$.

Furthermore, the total incoherent power spectrum is defined by
\begin{align}
F(k)=&\sum_{\alpha}\int d\omega S_\alpha^\text{incoherent}(\omega)\nonumber\\
=&\frac{4}{\pi}\int dxB_{k_1k_2}^*(x)B_{k_1k_2}(x).
\end{align}
It serves as a quantitative measure for the overall strength of correlations, and also provides a direct measure of the bound-state term. Under the assumption of a narrow bandwidth of incident photons, where the spectral width of the wave packet is significantly smaller than $\Gamma$, the wave packet approaches to a $\delta$ function. This implies that the incident photons have the equal frequency $k_1=k_2=k$. In this case, the expression for the total incoherent power spectrum can be simplified as
\begin{align}
F(k)=\frac{4\Gamma^3(1+\cos\vartheta)^3}{\pi\abs{k-\omega_0+i\Gamma^\prime}^4}=\frac{4}{\pi}\frac{\abs{B_{k_1k_2}(0)}^2}{\Gamma(1+\cos\vartheta)},
\end{align}
which is consistent with the single coupling point case (by replacing $\Gamma$ with $\Gamma/4$) when $\vartheta=0$~\cite{Y.L.L.Fang@pra2015}.
\begin{figure}[hbt]
\centering\includegraphics[width=8cm,keepaspectratio,clip]{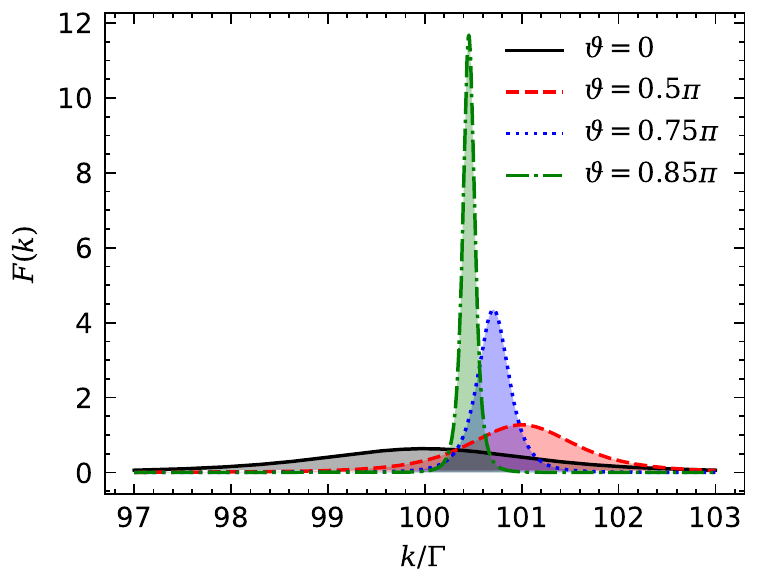}
\caption{The total incoherent power spectrum $F(k)$ as a function of the incident frequency $k$ with different values of the accumulated phase shift $\vartheta$. The other parameter is $\omega_0=100\Gamma$.}
\label{fig2}
\end{figure}

The total incoherent power spectrum $F(k)$ as a function of the incident frequency $k$ is shown in Fig.~\ref{fig2}. A large value of $F(k)$ indicates strong correlation effects, since the incoherent scattering originates from the correlation of the bound state. Therefore, the peak value of $F(k)$ corresponds to the strongest correlation, and $k_\text{peak}$ denotes the optimal incident frequency to obtain photon-photon correlations. It also shows that the peak value of $F(k)$ and $k_\text{peak}$ vary with the phase shift $\vartheta$. For example, when $\vartheta=0.85\pi$, it can be observed that the peak value of $F(k)$ increases by approximately an order of magnitude compared to its value at $\vartheta=0$. This implies a significant enhancement of photon-photon correlations. Physically, the position and width of the peak can be explained by the pole of the system, which corresponds to the zero of the denominator in the single-photon transmission or reflection amplitude $t(k)$ or $r(k)$. We denote the pole as $z=\tilde{\omega}-i\tilde{\Gamma}$, where the real part $\tilde{\omega}=\omega_0+\Gamma\sin\theta$ represents the eigenfrequency and $\tilde{\Gamma}=\Gamma(1+\cos\theta)$ denotes the effective decay rate. The position of the peak corresponds to the eigenfrequency  $\tilde{\omega}$, while its width is determined by $\tilde{\Gamma}$. Moreover, the peak value is given by $F_\text{peak}=4/\pi\tilde{\Gamma}$. In comparison to the single coupling point of a small atom, where the peak value is $8/\pi\Gamma$, it can be surpassed in the giant atom with $\vartheta\subset(2\pi/3, 5\pi/3)$. This indicates that the smaller effective decay rate can enhance photon-photon correlations, and it is adjustable through the change of the phase shift $\vartheta$ in the two-level giant atom system.

\subsection{Second-order correlation function}
\begin{figure}[hbt]
\centering\includegraphics[width=8cm,keepaspectratio,clip]{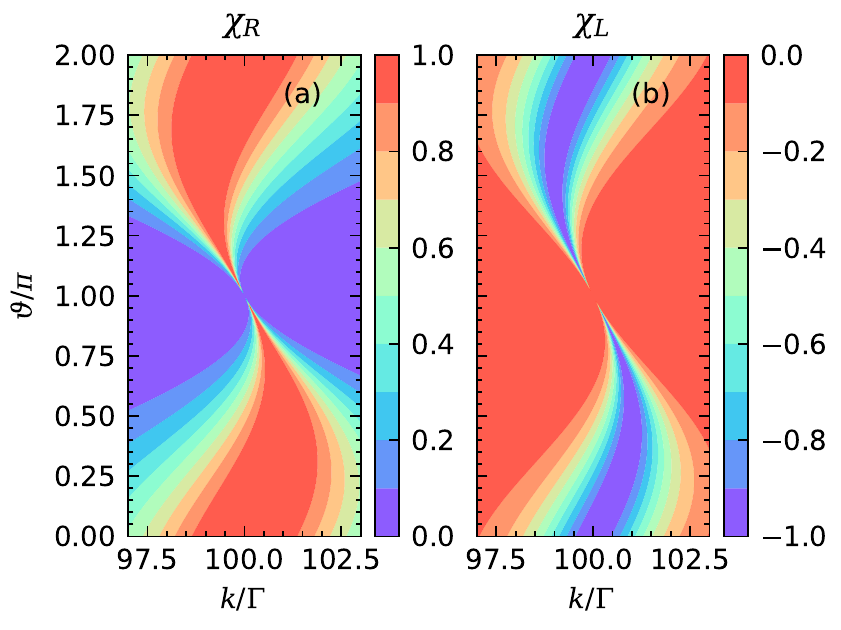}
\caption{The difference between the probability of two-photon detection and the independent single-photon detection when $x=0$, denoted as $\chi_R=2\pi^2\abs{f_{RR}(0)}^2-\abs{t_2(k)}^4$ for the transmitted field and $\chi_L=2\pi^2\abs{f_{LL}(0)}^2-\abs{r_1(k)}^4$ for the reflected field, as functions of the incident frequency $k$ and the accumulated phase shift $\vartheta$. The other parameter is $\omega_0=100\Gamma$.}
\label{fig3}
\end{figure}

Next, we utilize the second-order correlation function to demonstrate the spatial interaction between photons~\cite{loudon2000quantum}. The second-order correlation function of the transmitted and reflected fields ($x_1>d/2$, $x_2>d/2$ and $x=x_2-x_1$) are defined as follows:
\begin{align}
G_\alpha^{(2)}(x)=&\bra{\psi_2}\hat{a}^\dagger_\alpha(x_1)\hat{a}^\dagger_\alpha(x_2)\hat{a}_\alpha(x_2)\hat{a}_\alpha(x_1)\ket{\psi_2}\nonumber\\
&=2\abs{f_{\alpha\alpha}(x_1,x_2)}^2.
\end{align}
This correlation function represents the probability of detecting a photon at $x_2$ after detecting the first one at $x_1$. The expression is directly proportional to the rate at which two photons are transmitted or reflected, and is determined by the interference between the plane-wave term and the bound-state term. In order to briefly illustrate the effect of the bound state, we examine the difference between the probability of two-photon detection and the independent single-photon detection when $x=0$, under the condition that $k_1=k_2=k$, denoted as $\chi_R=2\pi^2\abs{f_{RR}(0)}^2-\abs{t_2(k)}^4$ for the transmitted field and $\chi_L=2\pi^2\abs{f_{LL}(0)}^2-\abs{r_1(k)}^4$ for the reflected field. If $\chi_R>0$, it indicates that the bound state enhances the transmission of two photons, resulting in a phenomenon known as photon-induced tunneling, which serves as a signature of photon bunching. Conversely, if $\chi_R<0$, it implies that the bound state can suppress the transmission of two photons, leading to photon blockade~\cite{H.Zheng@pra2012}. In Fig.~\ref{fig3}, it is shown that $\chi_R>0$, indicating that the transmitted photons are bunched, while $\chi_L<0$, suggesting that reflected photons are antibunched. Therefore, the statistical properties of photons are determined by the interference between the plane-wave and the bound-state terms.

\begin{figure}[hbt]
\centering\includegraphics[width=8cm,keepaspectratio,clip]{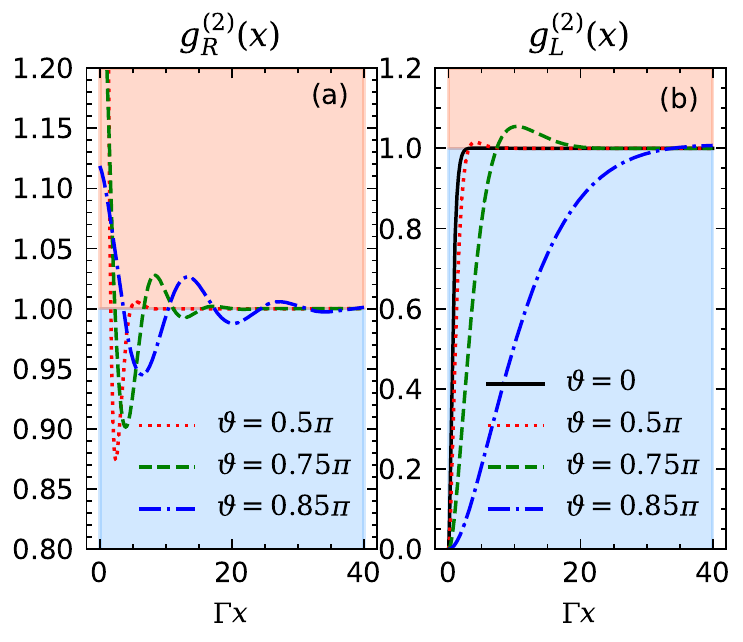}
\caption{The normalized second-order correlation functions, denoted as $g^{(2)}_R(x)$ for transmitted photons (a) and $g^{(2)}_L(x)$ for reflected photons (b), as a function of $x$ with different values of $\vartheta$. The other parameters are $\omega_0=100\Gamma$ and $k=\omega_0$.}
\label{fig4}
\end{figure}

It is also useful to introduce the normalized second-order correlation function~\cite{Y.Fang@EPJ2014,PhysRevA.107.023704,PhysRevA.101.053812}
\begin{align}
g_\alpha^{(2)}(x)=\frac{G^{(2)}_\alpha(x)}{\abs{_\alpha\langle x_1\ket{\phi_1(k_1)}_R}^2\abs{_\alpha\langle x_2\ket{\phi_1(k_2)}_R}^2}.
\end{align}
This function is normalized by the single-photon transmission and reflection probability. Furthermore, for incident photons with equal frequency, i.e., $k_1=k_2=k$, and by substituting the explicit expressions, the normalized second-order correlation functions can be simplified as
\begin{align}
g^{(2)}_R(x)&=\abs{1+\frac{\Gamma^2(1+\cos\vartheta)^2e^{i(k-\omega_0)\abs{x}-\Gamma^\prime\abs{x}}}{(k-\omega_0-\Gamma\sin\vartheta)^2}}^2,\nonumber\\
g^{(2)}_L(x)&=\abs{1-e^{i(k-\omega_0)\abs{x}-\Gamma^\prime\abs{x}}}^2.
\label{eqn:sec_correl}
\end{align}
These correlation functions are depicted in Fig.~\ref{fig4}. Here we choose the frequency of the input field to be resonant with the atomic transition frequency, i.e., $k=\omega_0$. As a result, the second-order correlation function is primarily determined by $\Gamma^\prime$. When $\vartheta=0$, the behavior of $g^{(2)}_{R/L}(x)$ resembles that of the single coupling point, which has been extensively investigated in theory as well as experiment with microwave photons~\cite{I.C.Hoi@prl2012}. The transmitted photons exhibit bunching behavior, while the reflected photons display anti-bunching. The correlations quickly reach the value of $1$ with little structure. Therefore, the initial value $g^{(2)}_{R/L}(0)$ makes a good prediction of the overall correlation nature. It should be noted here that at resonance, i.e., $k=\omega_0$, the single-photon transmission rate becomes zero, leading to a divergence in the normalized second-order correlation function of $g^{(2)}_R(x)$. In physics, $g^{(2)}_L(0)=0$ of the reflected field is due to that the two-level atom can only absorb and emit one photon at a time~\cite{I.C.Hoi@prl2012}.

However, when $\vartheta$ takes values of $0.5\pi$, $0.75\pi$, and $0.85\pi$, $g^{(2)}_{R}(x)$ exhibits oscillations between bunching and antibunching. Moreover, it takes a longer distance to reach the value of $1$. This behavior indicates that the photons become periodically organized in time and space. In contrast to the behavior observed in the case of $\vartheta=0$, where the initial correlation can predict whether the system generates bunching or antibunching of photons, the oscillatory feature presents a challenge in using the initial correlation as a reliable predictor. These features can be explained by the pole of the system, which appears in the exponential factor of Eq.~\eqref{eqn:sec_correl}. The oscillation arises from beating between the eigenfrequency and the incident frequency, and its duration is determined by the effective decay rate. Therefore, for $\vartheta=0$, the effective decay rate reaches the maximum value $2\Gamma$ and $g^{(2)}_{R/L}(x)$ rapidly reaches $1$. For $\vartheta=0.5\pi$, the effective decay rate becomes equal to $\Gamma$, resulting in an extended oscillation period. For $\vartheta=0.75\pi$ and $0.85\pi$, the effective decay rates further decrease, leading to even longer-lasting oscillations.

\section{The system of a three-level giant atom coupled to a 1D waveguide}
\label{sec3}
\begin{figure}[hbt]
\centering\includegraphics[width=8cm,keepaspectratio,clip]{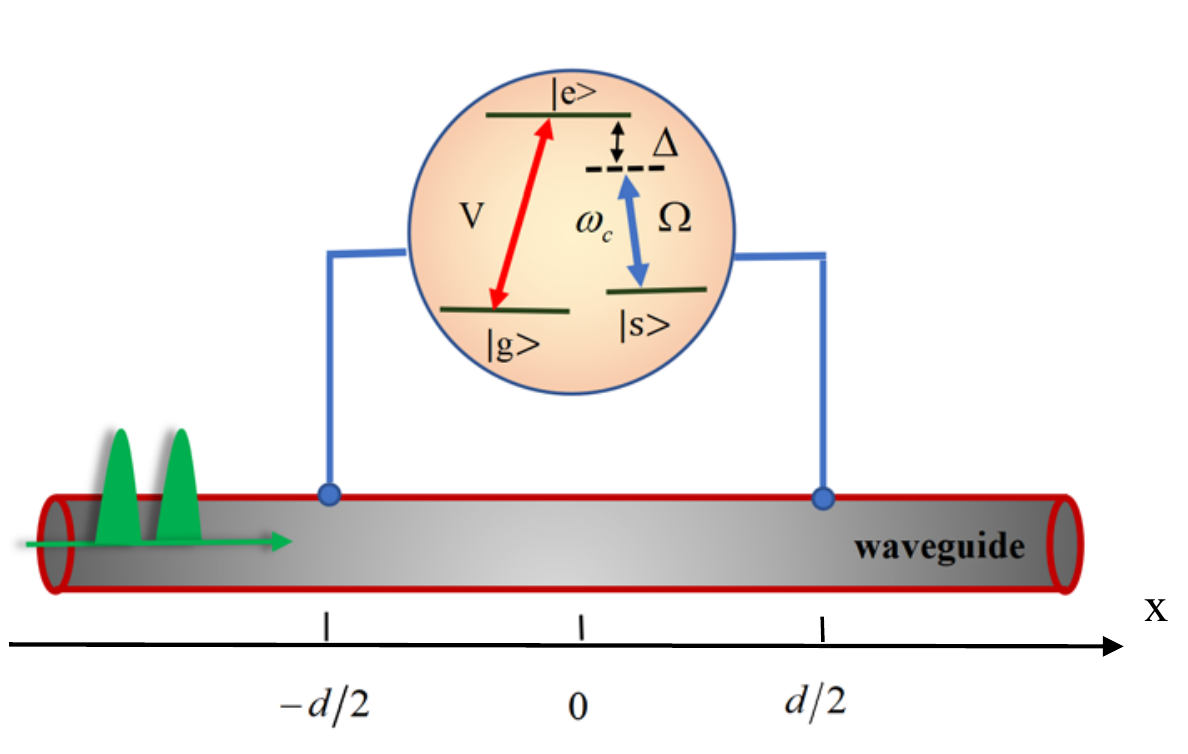}
\caption{Schematic illustration of a three-level giant atom side coupled to a 1D waveguide. The transition between the ground state $\ket{g}$ and the excited state $\ket{e}$ is coupled to waveguide modes positioned at $x=-d/2$ and $d/2$ with strength $V$. Additionally, the transition between the excited state $\ket{e}$ and the metastable state $\ket{s}$ is driven by a laser field of frequency $\omega_c$ with the Rabi frequency $\Omega$, and the detuning between the transition frequency and the laser frequency is $\Delta$.}
\label{fig5}
\end{figure}
We extend our investigation to explore the dynamics of correlated two-photon processes in a system consisting of a 1D waveguide side coupled to a three-level $\Lambda$-type giant atom, as shown in Fig.~\ref{fig5}. The configuration of a giant transmon coupled to a transmission line has been demonstrated in Ref.~\cite{A.M.Vadiraj@pra2021}, and its characteristic phenomenon, i.e., electromagnetically induced transparency (EIT), has been observed. For the single coupling point, it demonstrates that the second-order correlation of scattered photons can be tuned by adjusting the Rabi frequency of the control field~\cite{D.Roy@pra2014}. Now, our focus is specifically on investigating how the accumulated phase shift affects non-classical properties of scattered photons. The system under consideration is described by the Hamiltonian ($\hbar=1$ and the group velocity $\nu_g=1$)
\begin{align}
\hat{H}&=-i\int dx\left[\hat{a}^\dagger_R(x)\partial_x\hat{a}_R(x)-\hat{a}^\dagger_L(x)\partial_x\hat{a}_L(x)\right]\nonumber\\&+(\omega_0-i\gamma_e/2)\hat{\sigma}_{ee}+(\omega_0-\Delta)\hat{\sigma}_{ss}
+\frac{\Omega}{2}(\hat{\sigma}_{es}+\hat{\sigma}_{se})
\nonumber\\&+\frac{V}{2}\sum_{\alpha=R,L}\int dx M(x)\left[\hat{a}^\dagger_\alpha(x)\hat{\sigma}_{ge}+\hat{\sigma}_{eg}\hat{a}_\alpha(x)\right],
\end{align}
where $M(x)=\delta(x+d/2)+\delta(x-d/2)$ represents the positions of the coupling points, where $d$ is the separation between them. The parameter $\omega_0$ denotes the transition frequency of the atomic levels $\ket{g}\rightarrow\ket{e}$, and $\Delta$ represents the detuning between the transition frequency of $\ket{s}\rightarrow\ket{e}$ and the laser driving frequency $\omega_c$. It is worth noting that we consider the strong atom-waveguide coupling regime, where the interaction strength $\Gamma=V^2/2$ is significantly larger than the atomic spontaneous emission rate $\gamma_e$. The regime is realizable in experimental setups, and the atomic spontaneous dissipation rate $\gamma_e$ can be neglected. Within the single-excitation subspace, the eigenstate of the system can be written in the form
\begin{align}
\ket{\phi_1(k)}_\alpha=&\int dx\left[\phi_R^\alpha(k,x)\hat{a}^\dagger_R(x)+\phi_L^\alpha(k,x)\hat{a}_L^\dagger(x)\right]\ket{0,g}\nonumber\\&+u_e^\alpha(k)\ket{0,e}+u_s^\alpha(k)\ket{0,s}.
\end{align}
Here $\phi^\alpha_{R/L}(k,x)$ indicate the probability amplitudes of creating the right-moving and left-moving photon in real space for the $\alpha$-direction incoming photon with the wavevector $k$, respectively; $u_e^\alpha(k)$ is the excitation amplitude of the atomic level $\ket{e}$; $u_s^\alpha(k)$ is the excitation amplitude of the atomic level $\ket{s}$. These probability amplitudes can be determined by the Schr\"{o}dinger equation $\hat{H}\ket{\phi_1(k)}_\alpha=k\ket{\phi_1(k)}_\alpha$, which obey
\begin{align}
&(-i\partial_x-k)\phi_R^\alpha(k,x)+\sqrt{\frac{\Gamma}{2}}M(x)u_e^\alpha(k)=0,\nonumber\\
&(i\partial_x-k)\phi_L^\alpha(k,x)+\sqrt{\frac{\Gamma}{2}}M(x)u_e^\alpha(k)=0,\nonumber\\
&(\omega_0-k)u_e^\alpha(k)+\frac{\Omega}{2}u_s^\alpha(k)\nonumber\\&+\sqrt{\frac{\Gamma}{2}}\sum_{\alpha^\prime=R,L}\int dx M(x)\phi_{\alpha^\prime}^\alpha(k,x)=0,\nonumber\\
&(\omega_0-\Delta-k) u_s^\alpha(k)+\frac{\Omega}{2}u_e^\alpha(k)=0.
\end{align}
The solutions of these probability amplitudes take the following form
\begin{align}
\phi_R^R(k,x)=&\frac{e^{ikx}}{\sqrt{2\pi}}\big\{\theta(-d/2-x)+t_1(k)\big[\theta(x+d/2)\nonumber\\&-\theta(x-d/2)\big]+t_2(k)\theta(x-d/2)\big\},\nonumber\\
\phi_L^R(k,x)=&\frac{e^{-ikx}}{\sqrt{2\pi}}\big\{r_1(k)\theta(-d/2-x)\nonumber\\&+r_2(k)\left[\theta(x+d/2)-\theta(x-d/2)\right]\big\},\nonumber\\
\phi_R^L(k,x)=&\frac{e^{ikx}}{\sqrt{2\pi}}\big\{r_1(k)\theta(x-d/2)\nonumber\\&+r_2(k)\left[\theta(x+d/2)-\theta(x-d/2)\right]\big\},\nonumber\\
\phi_L^L(k,x)=&\frac{e^{-ikx}}{\sqrt{2\pi}}\big\{\theta(x-d/2)+t_1(k)\big[\theta(x+d/2)\nonumber\\&-\theta(x-d/2)\big]+t_2(k)\theta(-d/2-x)\big\},
\end{align}
where the coefficients are
\begin{align}
t_1(k)&=\frac{(\omega_0-\Delta-k)\left[\omega_0-k-i\frac{\Gamma}{2}\left(1+e^{i\vartheta}\right)\right]-\frac{\Omega^2}{4}}{(\omega_0-\Delta-k)\left[\omega_0-k-i\Gamma\left(1+e^{i\vartheta}\right)\right]-\frac{\Omega^2}{4}},\nonumber\\
t_2(k)&=\frac{(\omega_0-\Delta-k)\left(\omega_0-k+\Gamma\sin\vartheta\right)-\frac{\Omega^2}{4}}{(\omega_0-\Delta-k)\left[\omega_0-k-i\Gamma\left(1+e^{i\vartheta}\right)\right]-\frac{\Omega^2}{4}},\nonumber\\
r_1(k)&=\frac{i\Gamma(\omega_0-\Delta-k)(1+\cos\vartheta)}{(\omega_0-\Delta-k)\left[\omega_0-k-i\Gamma\left(1+e^{i\vartheta}\right)\right]-\frac{\Omega^2}{4}},\nonumber\\
r_2(k)&=\frac{i\Gamma/2(\omega_0-\Delta-k)(1+e^{i\vartheta})}{(\omega_0-\Delta-k)\left[\omega_0-k-i\Gamma\left(1+e^{i\vartheta}\right)\right]-\frac{\Omega^2}{4}},\nonumber\\
u_e^\alpha(k)&=\frac{-\sqrt{\Gamma/\pi}(\omega_0-\Delta-k)\cos\frac{\vartheta}{2}}{(\omega_0-\Delta-k)\left[\omega_0-k-i\Gamma\left(1+e^{i\vartheta}\right)\right]-\frac{\Omega^2}{4}},\nonumber\\
u_s^\alpha(k)&=\frac{\Omega/2\sqrt{\Gamma/\pi}\cos\frac{\vartheta}{2}}{(\omega_0-\Delta-k)\left[\omega_0-k-i\Gamma\left(1+e^{i\vartheta}\right)\right]-\frac{\Omega^2}{4}}.
\end{align}
We have made the assumption that the wavevector $k$ in the accumulated phase shift is a constant $k_0$ under the Markovian approximation, and replaced $k_0d$ with $\vartheta$.

In order to construct the wavefunction for two-photon scattering, the LS equation is employed by replacing the atomic operators with bosonic operators. To ensure compliance with level statistics, an additional on-site repulsion denoted as $U$ needs to be introduced and assumed to be infinite in the end. For a three-level atom, apart from the repulsion for each upper level, an additional term must be included to fully eliminate the double occupancy. Here the repulsion operator $\tilde{V}$ is
\begin{align}
\tilde{V}=\frac{U}{2}\left(\hat{b}_e^\dagger\hat{b}_e^\dagger\hat{b}_e\hat{b}_e+\hat{b}_s^\dagger\hat{b}_s^\dagger\hat{b}_s\hat{b}_s+2\hat{b}_e^\dagger\hat{b}_e\hat{b}_s^\dagger\hat{b}_s\right).
\end{align}
The coefficient of the last term is chosen for convenience, and any coefficient would be canceled out after taking $U\rightarrow\infty$~\cite{Y.Fang@physicaE2016}. By introducing an appropriate on-site interaction $\tilde{V}$, it becomes possible to calculate the two-photon wavefunction $\ket{\psi_2}$ following the preceding steps. As a result, in the limit  $U\rightarrow\infty$, the two-photon interacting eigenstate in the coordinate representation takes the form
\begin{align}
&_{\alpha_1^\prime\alpha_2^\prime}\langle x_1x_2\ket{\psi_2(k_1,k_2)}_{\alpha_1\alpha_2}=_{\alpha_1^\prime\alpha_2^\prime}\langle x_1x_2\ket{\phi_2(k_1,k_2)}_{\alpha_1\alpha_2}\nonumber\\&-\sum_{i,j=1}^3G^{\alpha_1^\prime\alpha_2^\prime}_i(x_1,x_2)(G^{-1})_{ij}\langle j\ket{\phi_2(k_1,k_2)}_{\alpha_1\alpha_2}.
\end{align}
Here, for simplicity, we denote $\ket{1}=\ket{d_ed_e}$, $\ket{2}=\ket{d_ed_s}$ and $\ket{3}=\ket{d_sd_s}$. The elements of the Green's functions are defined as follows:
\begin{align}
&G_i^{\alpha_1\alpha_2}(x_1,x_2)=_{\alpha_1\alpha_2}\langle x_1x_2\vert G^{R}(E)\ket{i}\nonumber\\&=\sum_{\alpha_1^\prime\alpha_2^\prime}\int dk_1dk_2\frac{_{\alpha_1\alpha_2}\langle x_1x_2\ket{\phi_2(k_1,k_2)}_{\alpha_1^\prime\alpha_2^\prime}\bra{\phi_2(k_1,k_2)}i\rangle}{E-k_1-k_2+i0^+},\nonumber\\
&G_{ij}=\bra{i}G^R(E)\ket{j}\nonumber\\&=\sum_{\alpha_1\alpha_2}\int dk_1dk_2\frac{\langle i\ket{\phi_2(k_1,k_2)}_{\alpha_1\alpha_2}\langle\phi_2(k_1,k_2)\ket{j}}{E-k_1-k_2+i0^+},\nonumber\\
&G^{-1}=\left(
         \begin{array}{ccc}
           G_{11} & G_{12} & G_{13} \\
           G_{21} & G_{22} & G_{23} \\
           G_{31} & G_{32} & G_{33} \\
         \end{array}
       \right)^{-1},
\end{align}
with the use of relations
\begin{align}
\langle 1\ket{\phi_2(k_1,k_2)}_{\alpha_1\alpha_2}=&u_e^{\alpha_1}(k_1)u_e^{\alpha_2}(k_2),\nonumber\\
    \langle2\ket{\phi_2(k_1,k_2)}_{\alpha_1\alpha_2}=&\frac{1}{2}\big[u_e^{\alpha_1}(k_1)u_s^{\alpha_2}(k_2)\nonumber\\&+u_e^{\alpha_1}(k_2)u_s^{\alpha_2}(k_1)\big],\nonumber\\
\langle3\ket{\phi_2(k_1,k_2)}_{\alpha_1\alpha_2}=&u_s^{\alpha_1}(k_1)u_s^{\alpha_2}(k_2),\nonumber\\
_{\alpha_1^\prime\alpha_2^\prime}\langle x_1x_2\ket{\phi_2(k_1,k_2)}_{\alpha_1\alpha_2}=&\frac{1}{2}\Big[\phi_{\alpha^\prime_1}^{\alpha_1}(k_1,x_1)\phi_{\alpha^\prime_2}^{\alpha_2}(k_2,x_2)\nonumber\\&+
\phi_{\alpha^\prime_1}^{\alpha_2}(k_2,x_1)\phi_{\alpha^\prime_2}^{\alpha_1}(k_1,x_2)\Big].
\end{align}

By employing the standard contour integral techniques to evaluate the double integral, we can obtain
\begin{widetext}
\begin{align}
G_{11}&=\frac{16\Gamma^2\cos^4\frac{\vartheta}{2}}{(\lambda_1-\lambda_2)^2}\left[\frac{(\lambda_1-\omega_0+\Delta)^4}{f_1(\lambda_1,\lambda_2)}
-\frac{(\lambda_1-\omega_0+\Delta)^2(\lambda_2-\omega_0+\Delta)^2}{f_2(\lambda_1,\lambda_2)}+\lambda_1\leftrightarrow\lambda_2\right],\nonumber\\
G_{22}&=\frac{4\Gamma^2\Omega^2\cos^4\frac{\vartheta}{2}}{(\lambda_1-\lambda_2)^2}\left[\frac{(\lambda_1-\omega_0+\Delta)^2}{f_1(\lambda_1,\lambda_2)}-\frac{(\lambda_1+\lambda_2-2\omega_0+2\Delta)^2}{4f_2(\lambda_1,\lambda_2)}+\lambda_1\leftrightarrow\lambda_2\right],\nonumber\\
G_{33}&=\frac{\Omega^4\Gamma^2\cos^4\frac{\vartheta}{2}}{(\lambda_1-\lambda_2)^2}\left[\frac{1}{f_1(\lambda_1,\lambda_2)}-\frac{1}{f_2(\lambda_1,\lambda_2)}+\lambda_1\leftrightarrow\lambda_2\right],\nonumber\\
G_{12}&=\frac{8\Omega\Gamma^2\cos^4\frac{\vartheta}{2}}{(\lambda_1-\lambda_2)^2}\left[\frac{(\lambda_1-\omega_0+\Delta)^3}{f_1(\lambda_1,\lambda_2)}-\frac{(\lambda_1-\omega_0+\Delta)^2(\lambda_2-\omega_0+\Delta)}{f_2(\lambda_1,\lambda_2)}+\lambda_1\leftrightarrow\lambda_2\right],\nonumber\\
G_{13}&=\frac{4\Omega^2\Gamma^2\cos^4\frac{\vartheta}{2}}{(\lambda_1-\lambda_2)^2}\left[\frac{(\lambda_1-\omega_0+\Delta)^2}{f_1(\lambda_1,\lambda_2)}-\frac{(\lambda_1-\omega_0+\Delta)(\lambda_2-\omega_0+\Delta)}{f_2(\lambda_1,\lambda_2)}+\lambda_1\leftrightarrow\lambda_2\right],\nonumber\\
G_{23}&=\frac{2\Omega^3\Gamma^2\cos^4\frac{\vartheta}{2}}{(\lambda_1-\lambda_2)^2}\left[\frac{(\lambda_1-\omega_0+\Delta)}{f_1(\lambda_1,\lambda_2)}-\frac{(\lambda_1-\omega_0+\Delta)}{f_2(\lambda_1,\lambda_2)}+\lambda_1\leftrightarrow\lambda_2\right],\nonumber\\
G_{21}&=G_{12}, \hspace{5pt} G_{31}=G_{13},\hspace{5pt}, G_{32}=G_{23},\nonumber\\
G_1^{RR}(x_1,x_2)&=\frac{2\Gamma\cos^2\frac{\vartheta}{2}}{\lambda_1-\lambda_2}\left[\frac{(\lambda_1-\omega_0+\Delta)(\lambda_1+\omega_0-\Delta-E)}{(2\lambda_1-E)(\lambda_1+\lambda_2-E)}e^{iEx_2-i\lambda_1t}-\lambda_1\leftrightarrow\lambda_2\right],\nonumber\\
G_2^{RR}(x_1,x_2)&=\frac{\Omega\Gamma\cos^2\frac{\vartheta}{2}}{2(\lambda_1-\lambda_2)}\left[\frac{2\omega_0-2\Delta-E}{(2\lambda_1-E)(\lambda_1+\lambda_2-E)}e^{iEx_2-i\lambda_1t}-\lambda_1\leftrightarrow\lambda_2\right],\nonumber\\
G_3^{RR}(x_1,x_2)&=-\frac{\Omega^2\Gamma\cos^2\frac{\vartheta}{2}}{2(\lambda_1-\lambda_2)}\left[\frac{1}{(2\lambda_1-E)(\lambda_1+\lambda_2-E)}e^{iEx_2-i\lambda_1t}-\lambda_1\leftrightarrow\lambda_2\right].
\end{align}
\end{widetext}
The expressions for $\lambda_1$ and $\lambda_2$ are given as
\begin{align}
\lambda_1=&\frac{1}{2}\Big[2\omega_0-\Delta-i\Gamma^\prime+\sqrt{(-\Delta+i\Gamma^\prime)^2+\Omega^2}\Big],\nonumber\\
\lambda_2=&\frac{1}{2}\Big[2\omega_0-\Delta-i\Gamma^\prime-\sqrt{(-\Delta+i\Gamma^\prime)^2+\Omega^2}\Big],
\label{eqn:lambda}
\end{align}
with $\Gamma^\prime=\Gamma(1+e^{i\vartheta})$. Furthermore, the functions $f_1(\lambda_1,\lambda_2)$ and $f_2(\lambda_1,\lambda_2)$ are defined as
\begin{align}
f_1(\lambda_1,\lambda_2)=&(2\lambda_1-E)(\lambda_1-\lambda_1^*)^2(\lambda_1-\lambda_2^*)^2,\nonumber\\
f_2(\lambda_1,\lambda_2)=&(E-\lambda_1-\lambda_2)(\lambda_1-\lambda_1^*)(\lambda_2-\lambda_2^*)\abs{\lambda_1-\lambda_2^*}^2.
\end{align}

During the contour integration, we utilize the conditions $x_1>d/2$ and $x_2>d/2$, and define $x=x_1-x_2$ and $x_c=(x_1+x_2)/2$. It can also be demonstrated that $G_i^{LL}(-x_1,-x_2)=G_i^{RL}(x_1,-x_2)=G_i^{LR}(-x_1,x_2)=G_i^{RR}(x_1,x_2)$ for $i=1,2,3$, owing to the parity symmetry. Finally, the expression for the two-photon interacting eigenstate becomes
\begin{align}
\ket{\psi_2(k_1,k_2)}_{RR}=&\int dx_1dx_2\Bigg[\frac{f_{RR}^{(\Lambda)}(x_1,x_2)}{\sqrt{2}}\hat{a}_R^\dagger(x_1)\hat{a}_R^\dagger(x_2)\nonumber\\
&+\frac{f_{LL}^{(\Lambda)}(x_1,x_2)}{\sqrt{2}}\hat{a}^\dagger_L(x_1)\hat{a}_L^\dagger(x_2)\nonumber\\
&+f_{RL}^{(\Lambda)}(x_1,x_2)\hat{a}_R^\dagger(x_1)\hat{a}_L^\dagger(x_2)\Bigg]\ket{0}.
\end{align}
The coefficients can be written as the sum of the plane-wave and bound-state terms
\begin{align}
f_{RR}^{(\Lambda)}(x_1,x_2)=&\frac{e^{iEx_c}}{\sqrt{2}\pi}\left[t_2(k_1)t_2(k_2)\cos\Delta_1x+B^{(2)}_{k_1k_2}(x)\right],\nonumber\\
f_{LL}^{(\Lambda)}(x_1,x_2)=&\frac{e^{-iEx_c}}{\sqrt{2}\pi}\left[r_1(k_1)r_1(k_2)\cos\Delta_1x+B^{(2)}_{k_1k_2}(x)\right],\nonumber\\
f_{RL}^{(\Lambda)}(x_1,x_2)=&\frac{e^{iEx/2}}{2\pi}\Big[t_2(k_1)r_1(k_2)e^{2i\Delta_1x_c}\nonumber\\&+r_1(k_1)t_2(k_2)e^{-2i\Delta_1x_c}+2B_{k_1k_2}^{(2)}(x_c)\Big].
\end{align}
The bound-state term $B^{(2)}_{k_1k_2}(x)$ at the resonance condition, i.e., $\Delta=0$, can be expressed as
\begin{align}
B_{k_1k_2}^{(2)}(x)=A_+e^{i(E/2-\gamma_+)\abs{x}}+A_-e^{i(E/2-\gamma_-)\abs{x}}.
\label{eqn:bound}
\end{align}
Here, $\gamma_\pm$ are the values of $\lambda_{1,2}$ in Eq.~\eqref{eqn:lambda} with $\Delta=0$, i.e.,
\begin{align}
\gamma_{\pm}=\omega_0-i\frac{\Gamma^\prime}{2}\pm\frac{\sqrt{\Omega^2-{\Gamma^\prime}^2}}{2},
\label{eqn:poles}
\end{align}
and the coefficients $A_\pm$ are
\begin{align}
A_\pm=&\frac{\Gamma^2(1+\cos\vartheta)^2}{2\sqrt{\Omega^2-{\Gamma^\prime}^2}}\Big[\pm(k_1+k_2-2\omega_0)\frac{\Omega^2}{4}\nonumber\\&
+(k_1-\omega_0)(k_2-\omega_0)(\sqrt{\Omega^2-{\Gamma^\prime}^2}\mp i\Gamma^\prime)
\Big]\nonumber\\&/\Big[(k_1-\gamma_+)(k_1-\gamma_-)(k_2-\gamma_+)(k_2-\gamma_-)\Big].
\end{align}
If $\Omega=0$, these coefficients can be reduced to the same form as those of the two-level giant atom derived above. In the following discussion, we also consider the narrow bandwidth limit for incident photons, where the frequencies of the incident photons are equal, i.e., $k_1=k_2=k$.

\subsection{Incoherent power spectrum}
\begin{figure}[hbt]
\centering\includegraphics[width=8cm,keepaspectratio,clip]{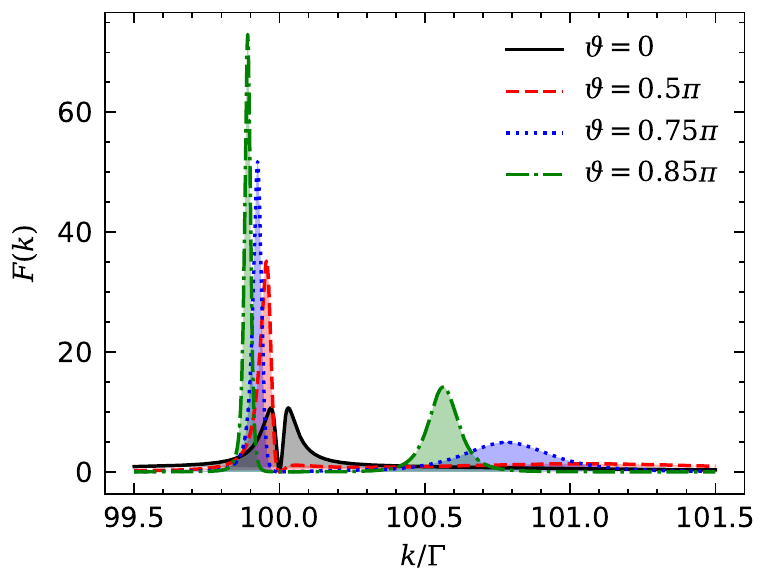}
\caption{The total incoherent power spectrum $F(k)$ of the three-level giant atom as a function of the incident frequency $k$ with different values of $\vartheta$. The other parameters are $\omega_0=100\Gamma$ and $\Omega=\Gamma/2$.}
\label{fig6}
\end{figure}

According to the previous discussion on the two-level giant atom system, the total incoherent power spectrum can be utilized as a metric for the photon-photon correlation because it originates from the correlated bound state. In the case of a three-level giant atom, the exact derivation of the total incoherent power spectrum $F(k)$ is provided in Appendix A, which is
\begin{align}
F(k)=\frac{8i}{\pi}\sum_{m,n=+,-}\frac{A_m^*A_n}{\gamma_m^*-\gamma_n}
\end{align}
The results are numerically shown in Fig.~\ref{fig6}. Notably, when the system reaches the perfect electromagnetically-induced transparency (EIT) window, i.e., $k=\omega_0$, the value of $F(k)$ becomes zero. This indicates that all photons are scattered coherently, and no bound state is formed in the wavefunction. This observation aligns with the expression of the bound-state term in Eq.~\eqref{eqn:bound}, which becomes zero when $k=\omega_0$. Thus, it is consistent with the notion that photon-photon correlation and incoherent scattering are related. Additionally, $F(k=\omega_0)=0$ at the perfect transparency remains unaffected by the phase shift $\vartheta$, and even the number of identical atoms coupled to the waveguide~\cite{Y.Fang@physicaE2016}. This phenomenon is commonly referred to as fluorescence quenching~\cite{P.Zhou@prl1996, E.Rephaeli@pra2011}.

A significant level of photon-photon correlation is indicated by a large value of $F(k)$. Figure~\ref{fig6} illustrates that the peak value of $F(k)$ for the three-level giant atom is considerably higher than that of the two-level giant atom. This suggests that the three-level giant atom generates photon-photon correlation more efficiently. Furthermore, when $\vartheta=0$, the shape of $F(k)$ exhibits symmetry and resembles that of the single-point coupling~\cite{Y.Fang@physicaE2016}. However, for other phase shifts such as $\vartheta=0.5\pi$, $0.75\pi$ and $0.85\pi$, the magnitude of incoherent scattering increases significantly, and the line shape changes from symmetry to asymmetry. This phenomenon can be explained by the presence of an asymmetric pole structure. The poles referred to here are $\gamma_{\pm}$ in Eq.~\eqref{eqn:poles}. When $\vartheta=0$, the real parts of the pole are symmetric with respect to $\omega_0$. However, for $\vartheta=0.5\pi$, $0.75\pi$ and $0.85\pi$, they become asymmetric. Additionally, the height of the peak is inversely proportional to its width, which is determined by the imaginary part of the pole. For example, when $\vartheta=0.5\pi$, the two poles are $\omega_0+1.03-0.97i$ and $\omega_0-0.032-0.03i$. As a result, the main peak is located at $\omega_0-0.032$ while the other peak is negligible due to its large imaginary part. Similarly, for $\vartheta=0.75\pi$, the two poles are $\omega_0+0.78-0.27i$ and $\omega_0-0.072-0.025i$, leading to the main peak being located at $\omega_0-0.072$. The height of the peak is higher due to the smaller imaginary part. At the same time, the other peak is located at $\omega_0+0.78$ with a smaller height due to its slightly larger imaginary part. Also, for $\vartheta=0.85\pi$, the two poles are $\omega_0+0.56-0.091i$ and $\omega_0-0.11-0.017i$, leading to the main peak being located at $\omega_0-0.11$ with a higher height. Simultaneously, the other peak located at $\omega_0+0.56$ also becomes evident due to its small imaginary part.

\begin{figure}[hbt]
\centering\includegraphics[width=8cm,keepaspectratio,clip]{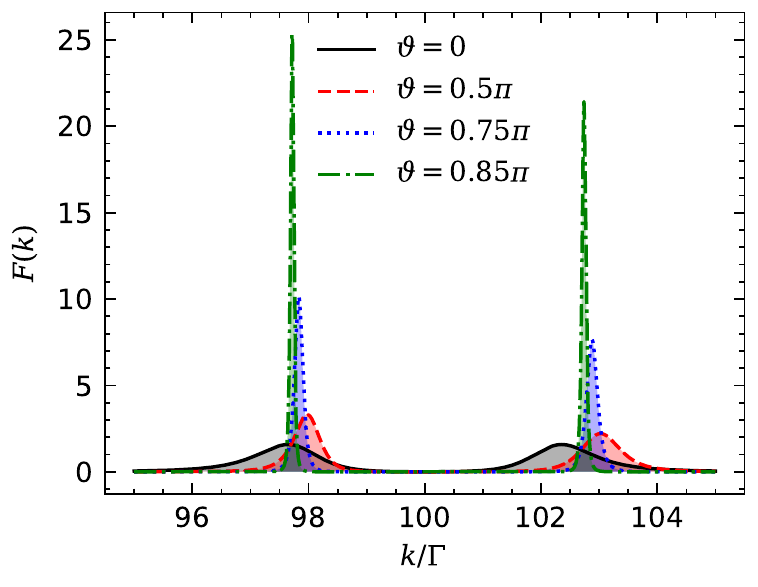}
\caption{The total incoherent power spectrum $F(k)$ of the three-level giant atom as a function of the incident frequency $k$ with different values of $\vartheta$. The other parameters are $\omega_0=100\Gamma$ and $\Omega=5\Gamma$.}
\label{fig7}
\end{figure}

When the Rabi frequency increases, the transparency window becomes wider. Within the transparency window, the scattered photons are not confined to the bound state, as shown in Fig.~\ref{fig7}. The peak values of $F(k)$ can be observed around $\omega_0\pm\frac{\Omega}{2}$, but their amplitudes decrease. Therefore, a strong Rabi frequency can hinder the photon-photon correlation and result in simpler and less structured effects. This characteristic may be advantageous in certain circumstances~\cite{D.Roy@pra2014}. The structure can also be explained by the poles $\gamma_\pm$, which approach $\omega_0\pm\frac{\Omega}{2}-i\frac{\Gamma^\prime}{2}$. The amplitudes become smaller due to the larger values of the imaginary parts.

\subsection{Second-order correlation function}
\begin{figure}[hbt]
\centering\includegraphics[width=8cm,keepaspectratio,clip]{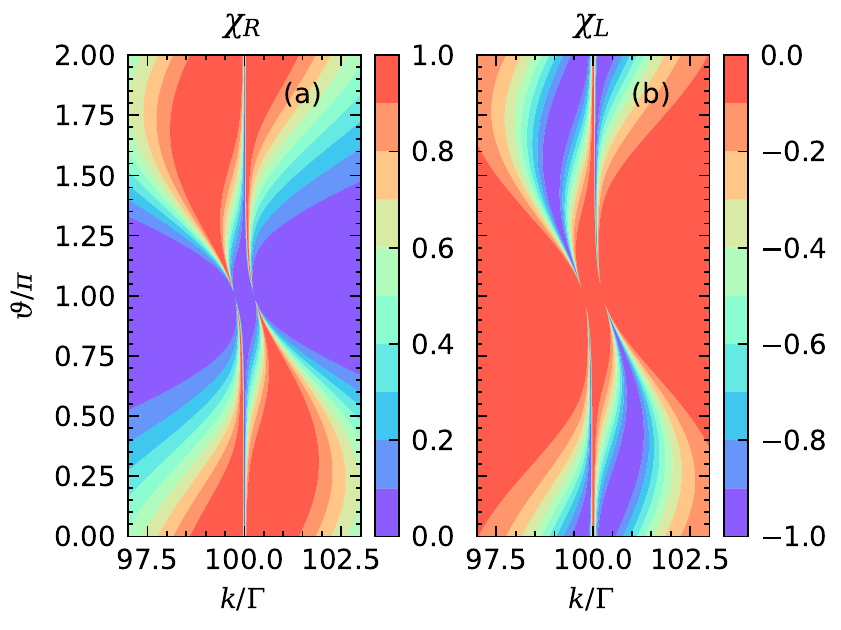}
\caption{The difference between the probability of the two-photon detection and the single-photon detection when $x=0$, denoted as $\chi_R=2\pi^2\abs{f_{RR}^{(\Lambda)}(0)}^2-\abs{t_2(k)}^4$ for the transmitted field (a) and $\chi_L=2\pi^2\abs{f_{LL}^{(\Lambda)}(0)}^2-\abs{r_1(k)}^4$ for the reflected field (b), as functions of the incident frequency $k$ and the accumulated phase shift $\vartheta$. The other parameters are $\omega_0=100\Gamma$ and $\Omega=\Gamma/2$.}
\label{fig8}
\end{figure}

Likewise, to briefly illustrate the effect of the bound state on the transmitted and reflected field ($x_1>d/2$, $x_2>d/2$, and $x=x_2-x_1$) in the system of the three-level giant atom, we first examine the difference between the probability of the two-photon detection and the single-photon detection when $x=0$. Under the condition of $k_1=k_2=k$, we denote $\chi_R=2\pi^2\abs{f_{RR}^{(\Lambda)}(0)}^2-\abs{t_2(k)}^4$ for the transmitted field and $\chi_L=2\pi^2\abs{f_{LL}^{(\Lambda)}(0)}^2-\abs{r_1(k)}^4$ for the reflected field. If $\chi_R>0$, it indicates that the bound state enhances the transmission of two photons, resulting in a phenomenon known as photon-induced tunneling, which serves as a signature of photon bunching. Conversely, if $\chi_R<0$, it implies that the bound state can suppress the transmission of two photons, leading to photon blockade. In Fig.~\ref{fig8}, it shows that $\chi_R>0$, indicating that the transmitted photons are bunched, while $\chi_L<0$, suggesting that the reflected photons are antibunched. Therefore, the statistical properties of photons are determined by the interference between the plane-wave and bound-state terms.

Similarly, the normalized second-order correlation functions are utilized to investigate the photon-photon correlation of the transmitted and reflected field in the system. Upon performing calculations, the normalized correlation functions can be expressed in the form
\begin{align}
g_R^{(2)}(x)=&\Bigg\vert1+\frac{A_+}{t_2(k_1)t_2(k_2)}e^{i(E/2-\gamma_+)\abs{x}}\nonumber\\&+\frac{A_-}{t_2(k_1)t_2(k_2)}e^{i(E/2-\gamma_-)\abs{x}}\Bigg\vert^2,\nonumber\\
g_L^{(2)}(x)=&\Bigg\vert1+\frac{A_+}{r_1(k_1)r_1(k_2)}e^{i(E/2-\gamma_+)\abs{x}}\nonumber\\&+\frac{A_-}{r_1(k_1)r_1(k_2)}e^{i(E/2-\gamma_-)\abs{x}}\Bigg\vert^2.
\end{align}
From the definition, it can be proven that at the perfect EIT window, where $k=\omega_0$, we have $g^{(2)}_{R/L}(x)=1$ due to the bound term in Eq.~\eqref{eqn:bound} being equal to zero. This implies that all of the photons are scattered coherently, without any structured correlation effects, which is consistent with the existence of a zero value for $F(k)$. It suggests that the statistical properties of the scattering photons can be adjusted by the control field in the three-level giant atom under the resonance condition. In the presence of the control field, the incident photons pass by the system coherently, thereby maintaining unchanged statistics. However, in the absence of the control field, the system effectively behaves as a two-level model, resulting in bunched transmitted photons and antibunched reflected photons. Actually, the generation of strong photon interactions indicated by the peak value of $F(k)$ appears slightly away from the EIT condition~\cite{D.Roy@pra2014}. Furthermore, it can be verified that when $x=0$, we have $g_L^{(2)}(0)=0$, indicating that the reflected photons are initially antibunched. As for the transmitted photons, the second-order correlation function at $x=0$ is given by
\begin{align}
g_R^{(2)}(0)=&\Bigg\vert 1+\Gamma^2(1+\cos\vartheta)^2\nonumber\\&\times\prod_{i=1}^2\frac{(k_i-\omega_0)}{(\omega_0-k_i)(\omega_0-k_i+\Gamma\sin\vartheta)-\frac{\Omega^2}{4}}\Bigg\vert^2.
\end{align}
It can be seen that when the incident photons have equal frequency, the transmitted photons are initially bunched. The analysis is consistent with the results presented in Fig.~\ref{fig8} and similar to that of the single coupling point~\cite{Y.Fang@physicaE2016}.
\begin{figure}[hbt]
\centering\includegraphics[width=8cm,keepaspectratio,clip]{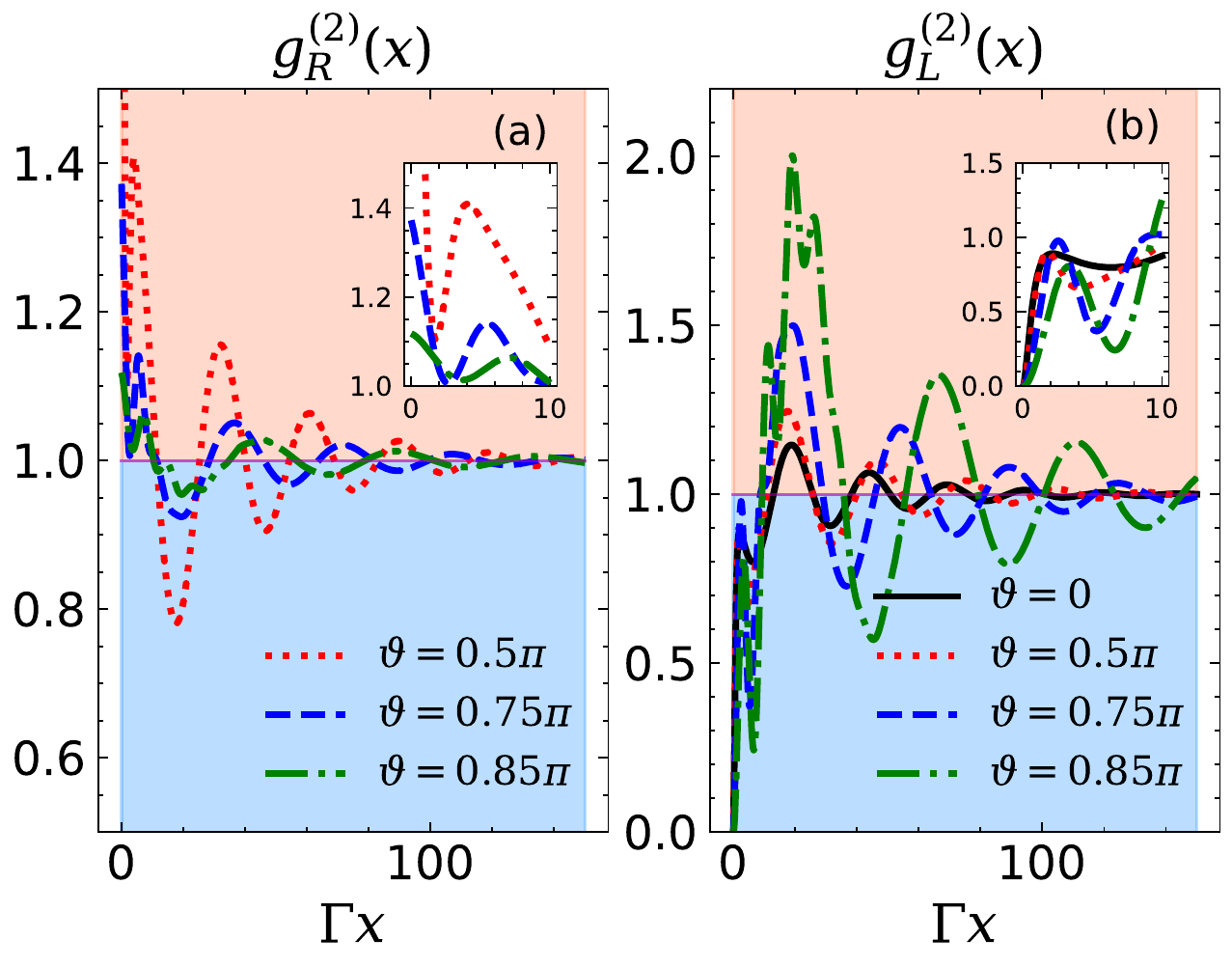}
\caption{The normalized second-order correlation functions, denoted as $g^{(2)}_R(x)$ for transmitted photons (a) and $g^{(2)}_L(x)$ for reflected photons (b), as a function of $x$ with different values of $\vartheta$. The other parameters are $\Omega=\Gamma/2$ and $\omega_0-k=\Omega/2$.}
\label{fig9}
\end{figure}

To illustrate the spatial correlation between photons, we have chosen $\omega_0-k=\Omega/2$ for the numerical plot of $g^{(2)}_{R/L}(x)$ in Fig.~\ref{fig9}. It should be noted that the initial value cannot be used to predict the overall photon-photon correlation due to the complex nature of $g_{R/L}^{(2)}(x)$ resulting from the beating between two eigenfrequencies and the incident frequency. For the given parameters in Fig.~\ref{fig9}, considering that $\abs{\text{Im}[\gamma_+]}\gg\abs{\text{Im}[\gamma_-]}$, the behavior of $g_{R/L}^{(2)}(x)$ at small values of $x$ is primarily determined by the term of $A_+$. It can be easily verified that $g_R^{(2)}(x)$ rapidly decreases within the scale of $\pi/(\text{Re}[\gamma_+]-k)$, while $g_L^{(2)}(x)$ quickly reaches a local peak at the same scale. On the other hand, for large values of $x$, the contribution of the term $A_+$ becomes negligible due to its fast decay. Instead, it is dominated by the beating of the term $A_-$ with the incident frequency. In addition to the enhanced photon-photon correlation compared to the two-level atom, the correlation lasts for a longer distance. The long decay distance can be characterized by the smaller value of the imaginary part of $\gamma_-$. When $\vartheta=0$, the decay time is on the scale of $8\Gamma/\Omega^2$, which is longer than that of two-level atom when employing the weak Rabi frequency $\Omega$. It should be noted here that the single-photon transmission rate becomes zero for $\vartheta=0$, resulting in a divergence in the normalized second-order correlation function of the transmitted field. Therefore, its evolution is not depicted in Fig.~\ref{fig9}. Moreover, for $\vartheta=0.5\pi$, $0.75\pi$, and $0.85\pi$, the imaginary parts become even smaller, indicating the possibility of achieving significantly long-distance correlations.

\section{Conclusions}
\label{conclusion}
In conclusion, we have employed the LS method to investigate the two-photon scattering processes involving two- and three-level giant atoms coupled to a 1D waveguide. We focus on effects of the accumulated phase shift acquired by photons as they travel between coupling points. The multiple coupling points of the giant atom give rise to interference effects that are absent in small atoms. Based on the analytical results for the total incoherent power spectra and second-order correlation functions of scattered photons, we have found that the accumulated phase shift can be utilized to enhance the strength and distance of photon-photon interactions. In the case of a two-level giant atom, photon-photon interactions are enhanced, and the evolution of the second-order correlation displays an oscillation between bunching and anti-bunching. Comparatively, the system of a three-level giant atom exhibits significantly increased photon-photon correlation, surpassing that of the two-level counterpart. The photon-photon interaction and the correlation distance can be further enhanced by tuning the accumulated phase shift between the two coupling points. These characteristics can be explained by analyzing the poles of the system.

\begin{acknowledgments}
This work was supported by the National Natural Science Foundation of China (under Grants No.1150403, 61505014,12174140.)
\end{acknowledgments}

\appendix
\section{Calculation of the incoherent power spectrum}
\subsection{The two-level giant atom}
The power spectrum or resonance fluorescence is defined as the Fourier transform of the first-order coherence function
\begin{align}
S_\alpha(\omega)=\int dt e^{-i\omega t}\bra{\psi_2}\hat{a}^\dagger_\alpha(x_0)\hat{a}_\alpha(x_0+t)\ket{\psi_2},
\end{align}
where $x_0$ is the position of a detector located far away from the scattering region. For example, considering $\alpha=R$, the first-order coherence in $\ket{\psi_2}$ is
\begin{align}
&\bra{\psi_2}\hat{a}_R^\dagger(x_0)\hat{a}_R(x_0+t)\ket{\psi_2}\nonumber\\&=2\int dx^\prime f_{RR}^*(x_0,x^\prime)f_{RR}(x_0+t,x^\prime).
\label{eqn:1st_coherence}
\end{align}
Under the condition of the narrow bandwidth for incident photons, they have equal frequency, i.e., $k_1=k_2=k=E/2$, and
\begin{align}
&f_{RR}^*(x_0,x^\prime)f_{RR}(x_0+t,x^\prime)=\nonumber\\&\frac{e^{iEt/2}}{2\pi^2}\Big[\abs{t_2(E/2)}^4+t_2^2(E/2)B_{k_1k_2}^*(x)\nonumber\\&
+t_2^{*2}(E/2)B_{k_1k_2}(x-t)+B_{k_1k_2}^*(x)B_{k_1k_2}(x-t)\Big],
\label{eqn:fRR_fRR}
\end{align}
with $x=x^\prime-x_0$. By substituting Eqs.~\eqref{eqn:1st_coherence} and~\eqref{eqn:fRR_fRR} into Eq.~\eqref{eqn:power_spectrum}, the power spectrum can be divided into two parts,
\begin{align}
S_R(\omega)=S_R^{\text{coh}}(\omega)+S_R^{\text{incoh}}(\omega),
\end{align}
where the coherent part contains terms proportional to $\delta(0)\delta(\omega-E/2)$ due to the use of delta-normalized plane waves. On the other hand, the incoherent part is related to the correlation of the bound-state term
\begin{align}
S_R^{\text{incoh}}(\omega)=\frac{1}{\pi^2}\iint dtdx e^{i(E/2-\omega)t}B_{k_1k_2}^*(x)B_{k_1k_2}(x-t).
\label{eqn:incoherent_spectrum}
\end{align}
After the integration, the incoherent power spectrum becomes
\begin{align}
S_R^{\text{incoh}}(\omega)=\frac{4}{\pi^2}\abs{\Upsilon(\omega)}^2,
\end{align}
where
\begin{align}
\Upsilon(\omega)=&\frac{\Gamma^2(1+\cos\vartheta)^2}{(E/2-\omega_0+i\Gamma^\prime)}\nonumber\\&\times\frac{1}{(E-\omega_0-\omega+i\Gamma^\prime)(\omega-\omega_0+i\Gamma^\prime)}.
\end{align}
The total incoherent power spectrum is obtained by summing the right- and left-moving incoherent power spectra,
\begin{align}
S^\text{incoh}(\omega)=S_R^{\text{incoh}}(\omega)+S_L^{\text{incoh}}(\omega).
\end{align}
Here it can be proven that $S_R^{\text{incoh}}(\omega)=S_L^{\text{incoh}}(\omega)$. The total incoherent power can be used to measure the overall strength of photon-photon correlations to show the non-classical effects. It also provides a direct measure of the bound-state term. The definition of the total incoherent power spectrum is
\begin{align}
F(k)=\int d\omega S^\text{incoh}(\omega)
=\frac{4}{\pi}\int dxB_{k_1k_2}^*(x)B_{k_1k_2}(x).
\end{align}
Via integrating over $\omega$, it becomes
\begin{align}
F(k)=\frac{4\Gamma^3(1+\cos\vartheta)^3}{\pi\abs{k-\omega_0+i\Gamma^\prime}^4}.
\end{align}

\subsection{The three-level giant atom}
The incoherent power spectrum in the system of the three-level giant atom can be calculated by following the same procedure as that of the two-level giant atom, which yields
\begin{align}
S_R^\text{incoh}(\omega)=\frac{1}{\pi^2}\abs{\sum_{m=+,-}\frac{A_m(E-2\gamma_m)}{(E-\omega-\gamma_m)(\omega-\gamma_m)}}^2,
\end{align}
where $\gamma_\pm=\pm\frac{1}{2}\sqrt{\Omega^2-\Gamma^{\prime2}}+\frac{i}{2}\Gamma^\prime$. By performing integration over the frequency $\omega$, the total incoherent power spectrum $F(k)=2\int d\omega S_R^\text{incoh}(\omega)$ becomes
\begin{align}
F(k)=\frac{8i}{\pi}\sum_{m,n=+,-}\frac{A_m^*A_n}{\gamma_m^*-\gamma_n}.
\end{align}

\bibliography{gutex}

\end{document}